# A focusing optical phased array for tissue interrogation with side-lobe suppression and simplified beam steering

PEDRAM HOSSEINI,[1] ALIREZA TABATABAEI MASHAYEKH,[1] PRACHI AGRAWAL,[1] YUNTIAN DING,[1] ALVARO MOSCOSO MÁRTIR,[1] REBECCA RODRIGO,[1] SANDRA JOHNEN,[2] FLORIAN MERGET,[1] AND JEREMY WITZENS[1*]

[1]*Institute of Integrated Photonics, RWTH Aachen University, Campus Boulevard 73, 52074 Aachen, Germany*
[2]*Department of Ophthalmology, University Hospital of RWTH Aachen, 52074 Aachen, Germany*
*\*jwitzens@iph.rwth-aachen.de*

**Abstract:** We implement an integrated multi-electrode array on a silicon-nitride-based photonic integrated circuit for *ex-vivo* retinal characterization via optical stimulation. The interrogation beam formers, based on curved grating emitters and optical phased arrays, are designed to achieve transverse focusing with spot sizes in the $1-2$ μm range to target single cells. The experimentally realized focusing optical phased arrays show suppressed side-lobes, with approximately 11.5% of the power in each side-lobe and ~60% in the main lobe, reducing unintentional cellular excitation. Additional design refinement enables further suppression of the side-lobes to a few percent of the total power. Additionally, we demonstrate a compact design of meandered thermal phase shifters implemented across the array that allow push-pull steering in the transverse direction as well as focusing and defocusing of the beam, with a total of only four control signals. Transverse angular steering by ±5.1° and axial translation of the focal spot by 204 μm are demonstrated with tuning currents below 50 mA, together with longitudinal angular steering by 4.26° obtained by means of wavelength tuning in a ±15 nm range centered on 525 nm.

## 1. Introduction

The combination of targeted optical excitation with electroretinography (ERG) is a powerful tool for understanding the specific functions of the individual cells in the different layers of the retina [1-3]. It is also useful in assessing the functional status of the retina, monitoring disease progression, developing therapies for specific retinal pathologies, and evaluating therapeutic effectiveness [4, 5]. Furthermore, optical probing can be a beneficial diagnostic tool for retinal degeneration, allowing for vision to be restored or enhanced through optogenetic treatments or by optimizing light wavelengths and intensities [6].

Developing precise optical or electrical stimulation and recording setups has been a primary focus from an engineering perspective [7, 8]. There have been many studies on enhancing electrical stimulation and recording using multi-electrode-arrays (MEAs) [9, 10]. However, electrical stimulation is associated with limitations, such as limited spatial resolution or cell-type selectivity, which can be overcome with optical stimulation methods [11]. In addition, combining optical stimulation with electrical recording of the retina provides enhanced synchronization across ganglion cells and results in a higher burst count [12]. Optical stimulation approaches have also demonstrated high spatial accessibility and precise local excitability at subcellular resolution [13]. Various studies have been published on state-of-the-art optical stimulation, based on a variety of approaches such as μLEDs [14], free-space optical systems [15], or micro-opto-electro-mechanical systems (MOEMS) [16]. However, each method has its own specific challenges and limitations, including limited resolution and significant heat dissipation in LED-based stimulators, the bulkiness and challenging alignment of free-space optical setups, and stray light in reflection-based MOEMS beam formers causing unintended stimulation.

Visible wavelength photonic integrated circuits (PICs) have been increasingly used in the life sciences for sensing and diagnostics applications [17, 18], microscopic imaging [19, 20], flow cytometry [21], and neurophotonic probes [22, 23]. Specifically, silicon nitride (SiN) PICs have shown promising results in biophotonics [24, 25] thanks to their small footprints, scalability, compatibility with standard complementary-metal-oxide semiconductor (CMOS) technology, and, most importantly, broad spectral transparency and low optical loss at visible wavelengths [26, 27]. Integrating optical beam formers with recording electrode systems on a PIC facilitates localized, high-resolution stimulation with high temporal precision [22, 28].

Free-space optics based focusing beam formers have allowed precise stimulation of retinal cells across different retinal layers [15, 29, 30]. There have been studies on PIC-based focusing emitters in other fields, such as optical trapping [31, 32], optically-enabled memory [33], or optical probes in optogenetics [23, 34]. The application of PICs to sub-cellular optical stimulation of biological tissues requires beam forming devices with sufficiently large apertures to achieve the required numerical aperture and thus sufficiently small spot sizes. In addition, beam steering is an essential capability for targeting specific cells once the tissue is affixed to the device. The field of view (FOV), that quantifies the achievable steering range, is thus another key metric for devices used in retinal studies [35, 36].

Optical phased arrays (OPAs) are planar beam-shaping and beam-steering devices that can be integrated at the chip scale and provide a large FOV. The beam, formed by the interference of the light emitted by the individual emitters, can be shaped or steered by manipulating the phase of coherent light in each of the channels of the array [37]. Many investigations have been carried out on these devices in PIC based optical systems [38-40], including, more recently, a number of biophotonic applications such as neural stimulation probes [41, 42].

The optical emitters used in OPAs can be based on end-fire or out-of-plane emission and consist of, for example, line beam emitters or grating emitters (GEs) [43]. The inherent wavelength dependency of the emission angle of GEs can be used to enable longitudinal beam-steering [43, 44], so that only transverse beam steering and, in the case of a focusing array, axial steering of the focal point have to be addressed with active phase shifters. Among these, thermo-optic phase shifters [45] are the least complex to implement, but also result in increased power consumption compared to micro-electromechanical system (MEMS) based devices [46] or phase shifters that rely on heterogeneous material integration [47]. One of the challenges addressed in this paper is how to design thermal phase shifters so as to minimize the number of independent control signals that have to be delivered to the chip, to facilitate routing, enable higher integration density, and simplify the resulting systems.

Due to their finite pitch, OPAs create a discretely sampled approximation of the targeted field emission, which results in the generation of side-lobes that may cause unwanted excitation of surrounding cells. Design methods for suppressing side-lobes in OPAs have previously been investigated. Spacing OPA channels at less than half of the emission wavelength ($\lambda/2$) results in single-lobe emission, a condition that is physically achievable in the silicon-photonic platform in the infrared spectrum. However, such close packing causes optical cross-talk among the OPA channels, necessitating complementary approaches such as amplitude apodization techniques [48, 49] or non-uniform array spacing [50, 51]. In the visible spectrum, decreasing the OPA pitch is further limited by the minimum feature size of the fabrication process, as a consequence of the smaller wavelengths. Another method, that has been implemented on an SiN PIC at visible wavelengths, involves filtering side-lobes using free-propagation in a slab, as shown in [52]. However, this technique results in the loss of the optical power confined in the filtered side-lobes and requires a large footprint for the slab, particularly in OPAs with a large number of channels. Suppressing side-lobes without losing the corresponding power in a compact device is the second challenge addressed in this paper.

We introduce an integrated setup for a MEA combined with a focusing PIC-based OPA, designed for *ex-vivo* characterization of the retina. We fabricate a focusing one-dimensional OPA (F-OPA) with 64 channels that features transverse focusing with a full width at half maximum (FWHM) of 1.27 µm at a 525 nm wavelength and a focal spot situated 200 µm above the surface of the chip, with ~60% of the total power concentrated in the main lobe. Tuning of the wavelength within ±15 nm of the center wavelength results in 4.26° of longitudinal beam steering. The side-lode suppression is further improved by an advanced design incorporating design techniques usually used in star couplers. The power in the main lobe is increased to ~80%, with only 1.6% of the power carried by each of the side-lobes and the remaining 17% distributed as stray light over the emission range.

Transverse steering and shifting of the focal point in the axial direction are achieved, respectively, by manipulating the phase across OPA channels linearly or parabolically with meandered thermal phase shifters that each only require a single control signal. A 16-channel F-OPA is fabricated that incorporates this steering concept with ±5.1° of transverse steering achieved by applying 50 mA to one or the other of the two complementary (push-pull) linear-phase-increment phase shifters implemented in the waveguide array. In addition, applying 50 mA to either a focusing or defocusing phase shifter results in a total focal spot displacement of 204 µm along the emission axis.

First, Section 2 presents the envisioned retinal characterization setup. Section 3 describes four different SiN-based F-OPAs with successive design improvements that suppress side-lobes with increasing effectiveness. Section 4 discusses the design and measurement of the thermal phase shifters, and finally, the obtained results are discussed and summarized. Limitations resulting from the waveguide coherence length as well as limitations of the measurement setup are discussed in the Appendix.

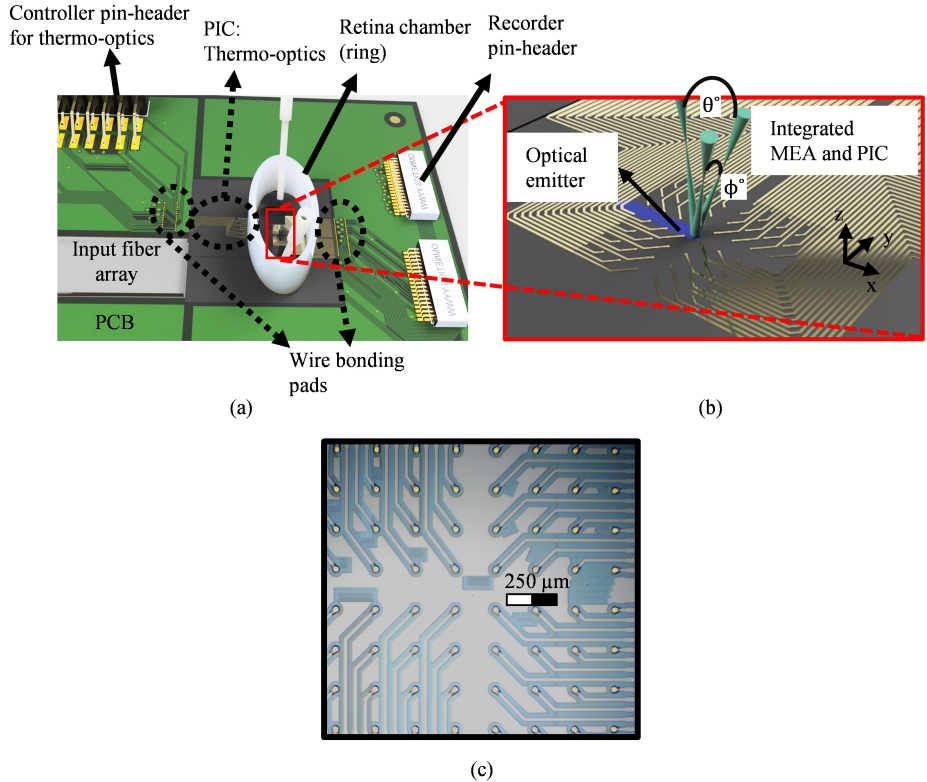

Fig. 1. (a) 3D model of the proposed setup: A PIC with optical emitters, a recording MEA, and thermo-optic phase shifters for programmable beam steering is mounted on a PCB for electrical interfacing via pin-headers to a control board for phase shifter configuration and an interface board for recording the neural activity. (b) Conceptual representation of a 64-channel gold-electrode MEA with an optical emitter enabling longitudinal (θ°) and transverse (φ°) beam steering as well as displacement of the focal point along the axis of the beam. (c) Fabricated PIC featuring the MEA and optical emitters located inside rectangular apertures defined in the shield electrode.

## 2. Multi-electrode array with focusing PIC-based optical emitter for retinal cell probing

Figure 1 shows a MEA for *ex-vivo* recording of retinal responses combined with SiN-based beam formers for optical probing. The MEA and beam formers have been combined on a PIC, whose micrograph is shown in Fig. 1(c) and which is mounted on and wire-bonded to a printed circuit board (PCB), as shown in Fig. 1(a). A polycarbonate ring attached to the PIC with a watertight biocompatible adhesive serves as a chamber for the retina. Oxygen-enriched nutrient medium can be flown into the chamber continuously, while used medium can be sucked out at the same time, to keep the tissue alive. The MEA and optical emitters are positioned inside, while thermal phase shifters used to dynamically steer the generated beams are located outside to prevent thermal damage to the tissue. The beams are generated through the rectangular apertures visible in Fig. 1(c) and are focused above the PIC at a programmable axial distance. Their longitudinal ($\theta$) and transverse ($\phi$) angle of propagation can also be steered, which enables addressing individual cells inside the tissue. The thermal phase shifters are connected via the PCB to a control board that allows the configuration of the beams, while the MEA is connected to an interface board that monitors the neuronal activity. It consists of 64 circular gold electrodes, each 500 nm thick including the adhesion layer, arranged on a 240 μm pitch, with each electrode having a diameter of 30 μm. Other optical components forming the light distribution network are covered by a metal shield electrode that minimizes unintended excitation caused by scattered light. The wavelength range of operation is centered on 525 nm to maximize the sensitivity of wild type (WT) mouse retina [53].

## 3. Optical beam formers

### 3.1 Grating emitter design on double stripe SiN platform and measurement setup

In this section, the detailed layer structure of the utilized SiN PIC platform and the base-designs of the GEs used to implement the beam formers are described, together with the test setup used to characterize the chips in the following sections.

The SiN PIC is based on the asymmetric double-stripe platform from LioniX International BV [26], with some customization applied to the layer thicknesses. Figure 2 shows its cross-sectional layer configuration. An 8 μm-thick silicon dioxide ($SiO_2$) bottom-cladding layer is grown on a silicon (Si) wafer through thermal oxidation. The waveguide core consists of two SiN layers, that are 27 nm (bottom) and 80 nm (top) thick and separated by a 50 nm $SiO_2$ interlayer. After patterning, these are covered by a 4 μm-thick $SiO_2$ top cladding layer. A 400-nm-thick gold metal interconnect layer is deposited over a 100-nm-thick Ta/Pt adhesion layer that also functions as a heater layer in regions where the gold has been selectively removed. These thin-film Ta/Pt heaters are employed in the thermo-optic phase shifters used to configure the beam. Finally, a 2 μm-thick $SiO_2$ passivation layer is deposited on top of the metal and removed on top of the electrodes of the MEA and interconnect pads for wire bonding.

The beam shapers described in this paper are all based on two baseline GE designs, GE1 and GE2, that are later adapted for focused beam emission or integrated into OPAs. As depicted in Fig. 2, in both cases the GE is formed by partially etching the top SiN layer, yielding better directionality and higher top-side emission than for fully etched devices. Key design parameters are the filling factor (FF), defined here as the proportion of the non-etched area inside the grating period ($\Lambda$), that are both constrained by fabrication driven design rules. In the first fabrication run, stepper lithography was used, with minimum feature sizes of 300 nm for top-layer etching and 400 nm for full etching. The second fabrication utilized electron-beam (e-beam) lithography for the top-layer etching inside the GEs, enabling smaller feature sizes down to 120 nm in non-etched areas and 250 nm in etched areas, limited in the latter case by the $SiO_2$ back-filling process.

To achieve single-mode waveguides (SMWs) at the target 525 nm wavelength with low propagation loss at the utilized transverse electric (TE) polarization, a fully etched, 400-nm-wide waveguide with an effective refractive index $n_{eff}$ of 1.608 was utilized for the interconnect network.

At the PIC input, an edge coupler with a vertical taper structure is used, transitioning adiabatically from a single-stripe waveguide core defined in the lower SiN layer, with a 400 nm width at the edge of the chip, to a double-stripe SMW. This taper was provided as a black-box, pre-designed component by the foundry.

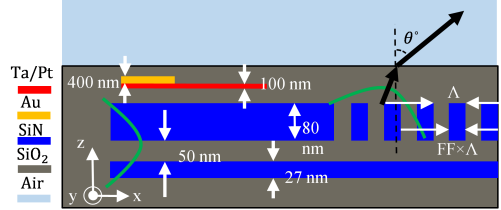

Fig. 2. SiN-PIC layer stack based on the LioniX International fabrication process.

The emission angle of the first diffraction order in the cladding ($\theta_{clad}$), relative to the normal to the chip surface, is determined by the phase-matching condition [54]

$$\sin(\theta_{clad}) = \frac{(1-FF)n_{eff1} + FF n_{eff2} - \frac{\lambda_0}{\Lambda}}{n_{clad}} \quad (1)$$

Here, $n_{eff1}$ and $n_{eff2}$ are effective refractive indices of the etched and non-etched waveguide sections, $n_{clad}$ is the refractive index of the SiO$_2$ cladding, and $\lambda_0$ is the vacuum wavelength.

Three-dimensional finite-difference time-domain (3D-FDTD) simulations were used to design efficient GEs. GE1 satisfies the design rules for the first fabrication run and has a period of 600 nm with a filling factor of 50%, while GE2 has a period of 420 nm and a 40% FF and requires e-beam lithography as applied to the second fabrication run. From these simulations, the electromagnetic fields in the x–y plane, at a height of 200 nm above the SiO$_2$-air interface (along the z-axis), were recorded. These fields were then two-dimensionally Fourier transformed to map them into k-space, with Fourier components resolved as a function of $k_x$ and $k_y$ as $F_{\alpha 0}(k_x, k_y)$, where α indexes the field component. To analyze far-field emission characteristics, the z-component of the Poynting vector, perpendicular to the $k_x$–$k_y$ plane, was calculated as

$$S_z(k_x, k_y) = \frac{1}{2}\text{Re}[E_x(k_x, k_y) \times H_y^*(k_x, k_y) - E_y(k_x, k_y) \times H_x^*(k_x, k_y)] \quad (2)$$

which is exemplarily visualized for GE2 in Fig. 3(a), in which the input waveguide has also been assumed to be broadened to 10 μm before reaching the first grate.

Integrating the 2D Poynting vector along the $k_y$-axis and subsequently calculating the propagation angle in air, $\theta_{air}$, using

$$\theta_{air} = \text{asin}(k_x/k_0) \quad (3)$$

and applying a change of variables $dk_x = k_0\cos(\theta_{air})d\theta_{air}$ yields the intensity distribution $S_\theta(\theta_{air}) = k_0\cos(\theta_{air}) \int S_z(k_x, k_y)dk_y$, with $k_0$ the magnitude of the free space k-vector. This angular emission profile is shown for both GE1 and GE2 in Fig. 3(b), from which it is apparent that GE1 has a large emission angle of $\theta_{air} = 43.5°$ for the first diffraction order and features a second, if much weaker, diffraction order. This motivated moving to e-beam lithography for the second run and designing GE2, which features a single diffraction order at $\theta_{air}$ = 16.4°.

To project the beam onto the x–y plane at various z-distances, the Fourier domain fields $F_{\alpha 0}(k_x, k_y)$ were first propagated by applying a phasor according to their k-vector component $k_z$, after which the inverse Fourier transforms were calculated according to

$$f_\alpha(x, y, z) = \mathcal{F}^{-1}[F_{\alpha 0}(k_x, k_y)e^{ik_z z}] \tag{4}$$

in which $k_z$ is given by

$$k_z = \sqrt{k_0^2 - k_x^2 - k_y^2} \tag{5}$$

To visualize the optical emission, the $u_z$-component of the Poynting vector, $S_{u_z}(x, y, z)$, was computed, where $u_z$ is oriented along the main direction of propagation of the beam. Figure 3(c) shows $S_{u_z}(x, z)$, after integration along the y-axis, for GE2.

Plotting the Poynting vector along the transverse and longitudinal cuts along the $u_x$–y and y–$u_z$ planes, with $u_x$ the normal to $u_z$ in the x–z plane, facilitates a comparison between simulations and measurements, since the beam is later recorded by a beam profiler with an optical axis oriented along the direction of propagation of the beam, as explained in the following.

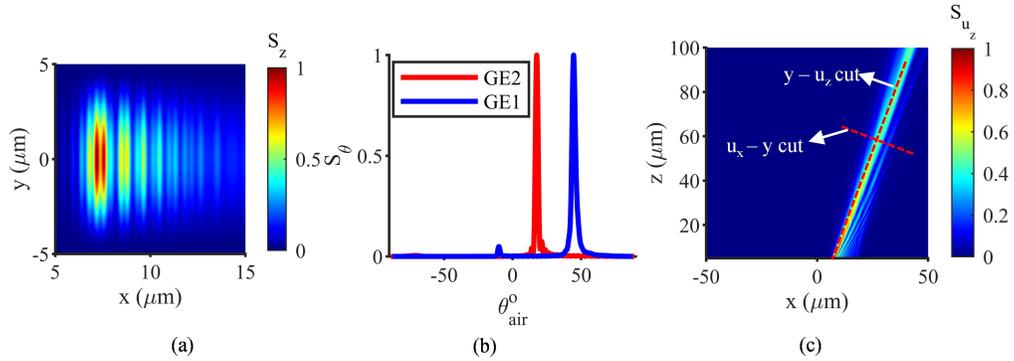

Fig. 3. (a) Near-field emission (z-component of the Poynting vector $S_z(x, y)$) from a 3D-FDTD simulation of GE2. (b) Angular emission pattern of GE1 and GE2. (c) Side-view (x–z plane) of the emission pattern of GE2 (component of the Poynting vector along the optical axis of the beam, $S_{u_z}(x, z)$).

The fabricated GEs are characterized using the setup shown in Fig. 4(a). A tunable laser with a center wavelength of 525 nm is coupled to the PIC via a lensed fiber (2 µm spot diameter, 10 µm focal distance) aligned precisely using a manual XYZ stage to an edge coupler on the chip, with a polarization controller ensuring that the TE mode is excited. A CMOS beam profiler equipped with a 4× C-mount beam expander (8 mm focal length, 0.922 µm effective pixel resolution after magnification) captures beam profiles at different distances from the chip.

In a first set of experiments aiming at characterizing the beam propagation angles, the beam profiler is first oriented with its optical axis along the normal of the chip surface (z-axis) without using the beam expander, at distances exceeding 1 mm from the PIC surface. By monitoring the position of the beam on the profiler as a function of its distance from the chip, the propagation angle is extracted for wavelengths spanning 510 nm to 540 nm, for further use in the following experiments. To generate experimental data that can be compared to beam profiles from simulations, the beam profiler is then oriented with a manual rotational stage such that its optical axis coincides with that of the beam and is mounted on a motorized XYZ stage that enables beam profiling in $u_x$–y planes at varying $u_z$.

The beam shapers implemented in this paper aim at providing dynamic beam steering in both the longitudinal (θ) and transverse (φ) directions (Fig. 1(b)). The longitudinal steering is obtained via wavelength tuning and discussed here, while the transverse and axial beam steering via thermal phase shifters are discussed in Section 3.3.

Equation 1, relates the emission angle of the GEs to their operating wavelengths, which enables steering in the θ-direction [55]. Taking a grating emitter with a width of 10 μm, 30 grates and the grating parameters of GE2, wavelength tuning between 510 nm and 540 nm results in θ being scanned in a range of 4.85° in the simulations. This is close to the measurement results which are overlaid in Fig. 4(b) and feature a 4.26° steering angle. Corresponding data for GE1 has been reported in [56].

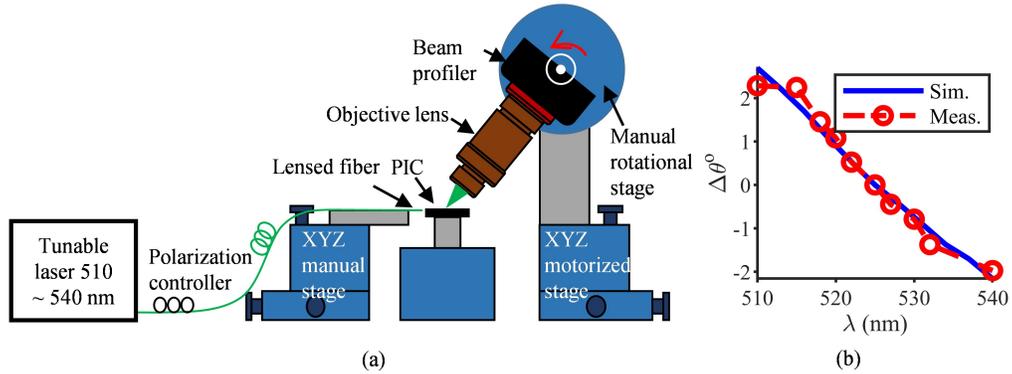

Fig. 4. (a) Experimental setup for the far-field characterization of the PIC-based beam formers. (b) Longitudinal beam steering measured for GE2 in a 510 nm to 540 nm wavelength range, showing good consistency between simulation and experiment.

As mentioned before, one goal is to design an optical emitter with sufficient focusing above the chip surface to enable subcellular resolution. A curved grating structure embedded in a straight slab waveguide is first designed and shown in Fig. 5(a), which focuses the beam along the transverse y-direction at a target distance of ~156 μm above the chip. The focusing grating emitter (FGE) is based on the grating structure of GE1 and designed with a radius of curvature of 180 μm for the first grate and a width of 123 μm along the y-direction. A 1 mm long taper expands the beam between the SMW and the grating, designed using 2.5 FDTD simulations. A design trade-off is struck between a small footprint and letting the field expand across most of the taper width. Ideally, the phase front at the grating would be flat, with the field corresponding to the ground mode of the local taper cross-section and the curvature of the grates serving to focus the beam out of the plane of the chip. With a 1 mm length, the taper is, however, not fully adiabatic, resulting in some residual phase front curvature inside it that pushes out the focal spot somewhat further out from the surface of the chip and reduces the effective numerical aperture of the grating, as the field does not cover its entire span.

This configuration produces a focal spot with a FWHM in the transverse y-direction of ~1.44 μm at a distance of 232 μm measured from the PIC surface along the propagation direction. Thereby, the designed FGE offers a higher resolution than reported in [34]. Figure 5(c) visualizes the simulated Poynting vector component $S_{u_z}(y, u_z)$ in the y–$u_z$ plane, that can be directly compared to the experimental data visualized in Fig. 5(d). The experimental data is obtained by sequentially recording the intensity in different $u_x$–y planes, at varying $u_z$, and stitching the data back together. In both plots, the recorded Poynting vector component / intensity has been integrated along $u_x$ prior to plotting in the y–$u_z$ plane. Simulation and experiment show good consistency, both in terms of the position and the width of the focal spot (see Table 1).

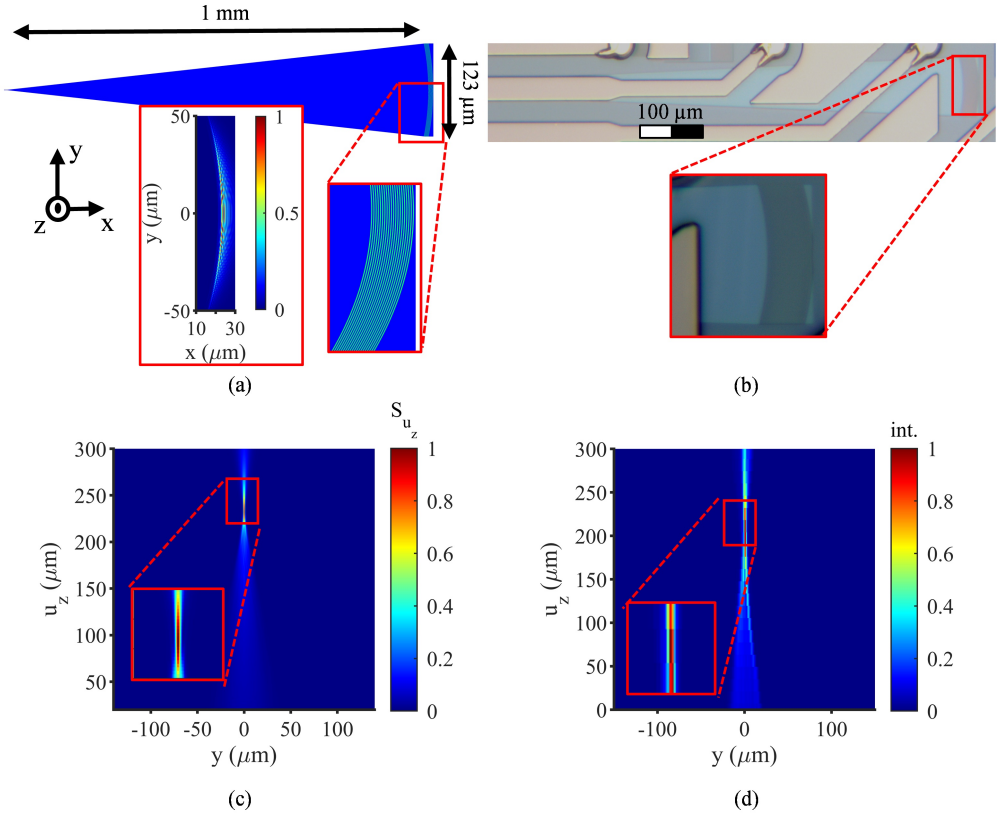

Fig. 5. Focusing grating emitter (FGE). (a) Layout with an inset showing 3D-FDTD simulation of $S_z(x, y)$ in the near field. (b) Micrograph of the fabricated structure. $y$–$u_z$ view of (c) the simulated beam ($S_{u_z}(y, u_z)$) and (d) the experimentally recorded intensity.

Thus far, SiN-PIC-based GEs have been demonstrated with subcellular resolution in the transverse direction and dynamic longitudinal beam steering. To achieve full spatial coverage, transverse steering ($\phi$) and axial focal spot movement are required; therefore, OPAs, discussed in the following sections, serve as alternatives to single-input GEs.

### 3.2 Design and simulation of focusing OPA

Figure 6(a) illustrates a PIC-based 64-channel one-dimensional (1D) OPA with focusing capabilities (F-OPA), as investigated in this study.

It consists of a 1D GE array, which can be designed conventionally from spatially separated GEs as depicted in Fig. 6(c). Alternatively, the fields provided by the interconnect waveguides can be combined in a single slab featuring a uniform grating structure, as shown in Fig. 6(d). The GE array in Fig. 6(c) is characterized by its spacing and the pitch, denoted as d. As we shall see in the following, the choice of configuration has important consequences on the magnitude of parasitic side-lobe formation.

To ensure coherent light coupling into the OPA, the $2^n$ emitters are supplied with light by a single input fiber, which is coupled to the PIC through an edge coupler. The light then propagates through a network with n levels of 1×2 multi-mode interferometers (MMIs), that distributes the optical power evenly among all the GEs.

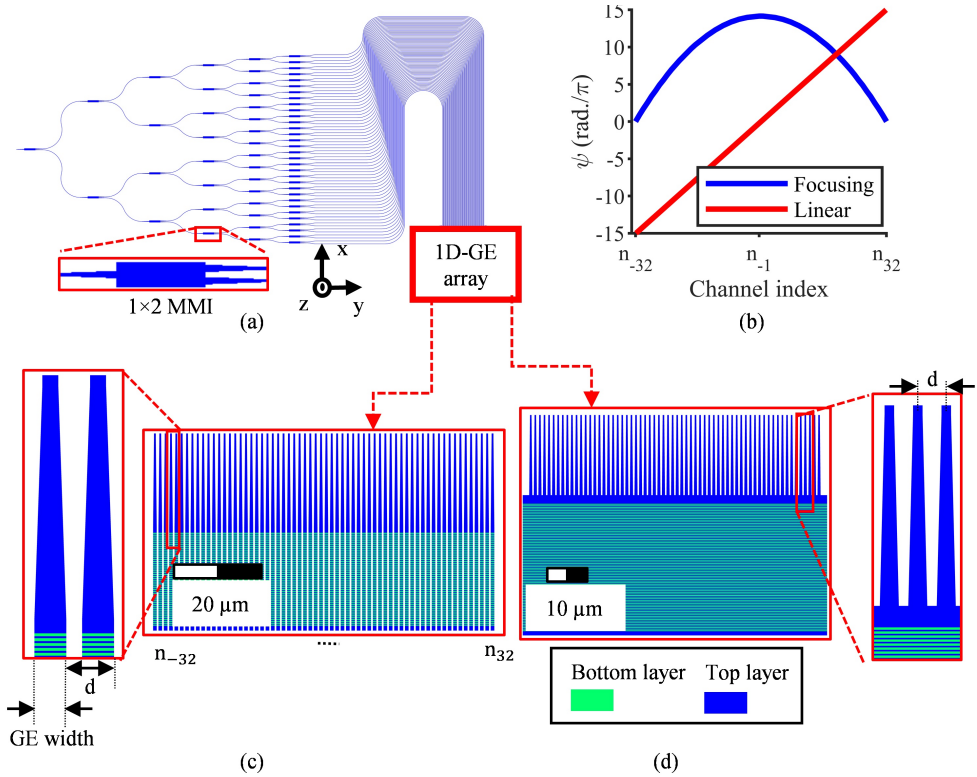

Fig. 6. (a) Exemplary layout of a 64-channel focusing 1D-OPA, including a 6-level MMI-based optical power distribution network and GE-array. (b) Illustration of parabolic and linear phase offsets between OPA channels resulting in beam focusing and steering. Layout details for (c) F-OPA1, a 1D array of discrete GEs (pitch d = 1.2 μm, emitter width = 0.8 μm), and (d) F-OPA2, a 1D array of discrete waveguides connected to a single slab GE.

As illustrated in Fig. 6(b), introducing a linear or parabolic phase dependency across the GEs enables configuration of the emitted beam, with the first resulting in steering in the transverse (y-)direction and the second resulting in focusing, also in the transverse direction. In the following, the parabolic phase difference is applied passively, through built-in delay lines implemented by U-shaped routing of the interconnect waveguides (as shown in Fig. 6(a)), resulting in a baseline focal point position above the chip surface, and actively, by using phase shifters as discussed in Section 3.3, which enable further tuning of the focal spot position. Similarly, transverse beam steering will also be enabled with phase shifters.

Introducing a linear phase offset with an increment Δψ across adjacent GEs results in an emission angle ϕ in the transverse direction described by

$$\sin \phi = \frac{\lambda_0 \Delta \psi}{2\pi d} \qquad (6)$$

Due to the discrete nature of the emitters or of the waveguide channels inside the OPAs, higher-order diffraction occurs, creating side-lobes in addition to the fundamental one. The angular separation Δϕ to the main lobe can be calculated as

$$\sin(\phi + \Delta\phi) - \sin(\phi) = \frac{\lambda_0}{d} \qquad (7)$$

The 1D-GE arrays depicted in Figs. 6(c) and 6(d) both have a pitch of 1.2 μm and incorporate GEs matching the parameters of GE2. We refer to them as F-OPA1 and F-OPA2, respectively, with cumulative widths of 76.4 µm and 79.8 µm. In F-OPA1, SMWs transition into multi-mode waveguides (MMWs), each connected to a discrete GE arranged in an array, resembling a scaled up version of the 16-channel focusing OPA described in [56]. Conversely, in F-OPA2, all SMWs transition into a single slab waveguide with an embedded continuous grating structure.

Both systems utilize tapers designed to achieve adiabatic transitions to the final MMW width over a 20 µm length. Due to minimum feature size constraints, MMWs maintain a 0.4 µm separation gap before the onset of the GEs or slab. The side walls of the waveguides as well as the onset of the slab region in F-OPA2 are defined by full etches extending through both SiN layers.

The optical distribution network includes 1×2 MMIs measuring 20.1 µm in length and 3.5 µm in width. Additionally, a bend radius of 35 µm is utilized in the interconnect waveguides, resulting in negligible bending losses.

To achieve static beam focusing, delay lines are implemented along the waveguide channels that generate a parabolic phase distribution, that corresponds to a maximum phase delay at central waveguides and gradually reduces to zero towards the outer ones. The channel-dependent parabolic phase distribution is defined by

$$\psi_{ch}(p) = 4\psi_M \frac{p(N-1-p)}{N(N-2)} \quad (8)$$

where $p \in [0, N-1]$ indexes the channel number, $\psi_{ch}(p)$ represents the phase delay of channel p, $\psi_M$ is the maximum phase delay applied to the two central channels, and N is the total number of channels. The corresponding waveguide delay lengths $L_\phi$ required to induce these phase offsets are calculated as

$$L_\phi(p) = \frac{\lambda_0 \psi_{ch}(p)}{2\pi n_{eff}} \quad (9)$$

The waveguide lengths are calculated accordingly to obtain a parabolic phase delay with a maximum phase $\psi_M$ set to 45 radians. In addition, all structures (straight waveguides, bends, MMIs, GEs) are designed to have positions and dimensions snapped to the design grid of the mask making process, to increase the reproducibility of the devices.

To reduce the computational time required to model the OPA, we start with a 3D-FDTD simulation modeling an array of five channels that involves launching a single mode into the central waveguide and two adjacent waveguides on either side of it in which light is not individually launched, but to which light can be cross-coupled from the central one. These are included in the simulation to consider cross-coupling effects during the SMW-to-MMW transitions. The number of waveguides on either side is sufficient here, as the cross-coupling effects are too weak to reach waveguides that are further out.

As previously, the electromagnetic fields are recorded in the x–y plane right above the PIC. Figures 7(a) and 7(d) present the calculated near-field distributions, represented as the z-component of the Poynting vector, $S_z(x, y)$, for F-OPA1 and F-OPA2, respectively. The overall fields above the surface of the chip are then reconstructed by summing the contributions for each source waveguide. In other words, the recorded fields are translated by an integer multiple of d along the y-direction to recenter them on the corresponding waveguide, multiplied by a phasor applying the designed-in parabolic phase, and summed together for all the sources, which effectively amounts to a convolution. The translation and phase shift operations can be applied together according to

$$F_{\alpha p}(k_x, k_y) = F_{\alpha 0}(k_x, k_y) e^{i(\psi_p - k_y y_p)} \quad (10)$$

where $\psi_p$ is the phase delay applied to channel p and $y_p$ is the center position, along the y-axis, of the GE of index p, respectively of the corresponding MMW. The additional phase $-k_y y_p$ translates the envelope of the field profile to the position of the corresponding waveguide.

Figures 7(b) and 7(e) illustrate the normalized intensity $S_{u_z}(u_x, y)$ in the focal plane on a logarithmic scale for F-OPA1 and F-OPA2, respectively. The maximum intensity is normalized to 1 (0 dB). Both devices exhibit two side-lobes in addition to the main one. These side-lobes are located 86.8 µm away from the main lobe along the y-axis, with focal spots at distances of 190 µm as measured along $u_z$.

To quantify the power distribution, $S_{u_z}(u_x, y)$ is integrated across rectangular monitors centered over each of the lobes. These monitors span the full simulation region along the $u_x$-axis, with a width along the y-axis equal to twice the mode field diameter (MFD) obtained from the Gaussian fit of each lobe (Fig. 7(b)), see Appendix A.2 for individual fits. Power falling outside the monitors defined for each lobe is considered stray light. The power contained in the main lobe is 56.65% for F-OPA2, which is higher than the 53.78% observed for F-OPA1. Figures 7(c) and 7(f) show $S_{u_z}(y, u_z)$ in the y–$u_z$ plane, further illustrating the emission pattern.

Replacing discrete GEs by a slab waveguide with a continuous grating emitter structure, similarly to what is presented in [41], is seen here to reduce the power contained in the side-lobes. While this initial improvement is marginal, it indicates a promising path for further side-lobe suppression that is explored in the following.

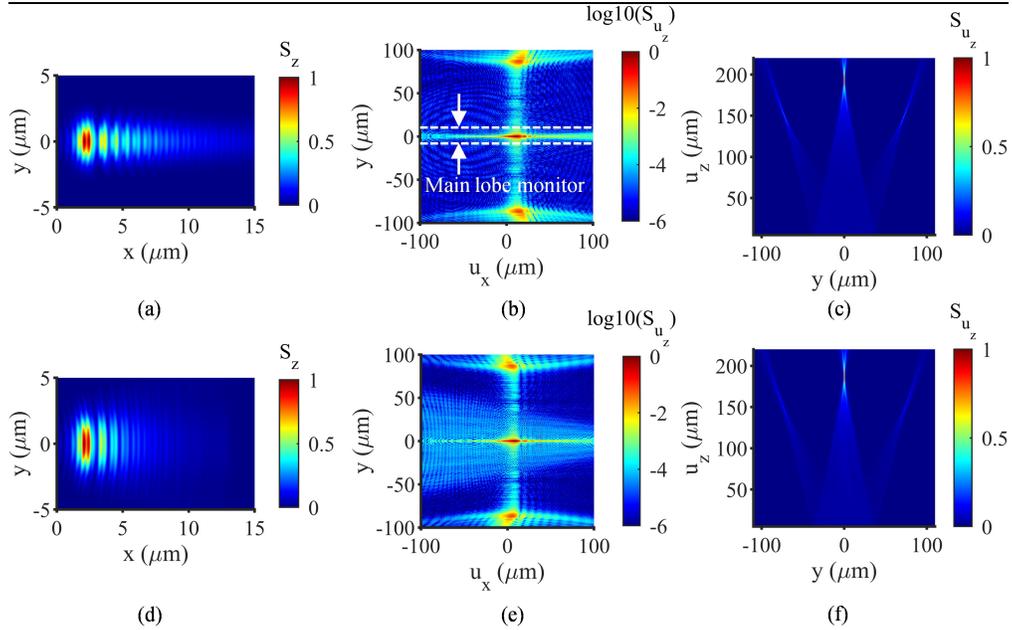

Fig. 7. Simulation results for F-OPA1 (top row) and F-OPA2 (bottom row). (a), (d) Near-field distribution (z-component of the Poynting vector, $S_z(x, y)$) from a 3D-FDTD simulation of an OPA with 5-channels and a single mode launched into the center one. Distribution of the axial Poynting vector component ($S_{u_z}$) in (b), (e) the focal plane of the OPA (logarithmic scale) and (c), (f) the y–$u_z$ plane, obtained after convolution and propagation of the field.

### 3.2.3 Focusing optical phased arrays with enhanced side-lobe suppression

In addition to reducing the power contained in the main lobe of the presented F-OPAs, side-lobes may unintentionally excite other cells located on their propagation path and are thus undesirable. To address this, we investigate two F-OPA designs that further reduce side-lobe emission. For this, we can follow the same path as used in star couplers and arrayed waveguide gratings to suppress higher diffraction orders [57, 58]. Important criteria for suppressing side-lobes inside the slab region of properly designed star couplers are that (i) the waveguides in the arrays at its interfaces need to be strongly coupled to each other and that (ii) dummy waveguides have to be appended to the edges of the array to mimic an infinite array from the perspective of the electromagnetic field.

To understand why, one can consider that an array of waveguides with a pitch larger than half the wavelength ($d \gg \lambda/2$) forms a periodic structure whose super-modes are Bloch modes along the direction of the array (y). In the initial array of uncoupled SMWs, the incoming light can straightforwardly be decomposed into a basis consisting, for each basis vector, in light being excited in a single SMW, or equivalently, and more intuitively for what is to follow, in a basis in which each basis vector consists in equal power being launched in each of the waveguides, but with a waveguide specific phase given by $\psi_p = 2\pi pq/N$, where $q \in [0, N-1]$ is the index of the basis vector. Inside the slab, after the SMW to MMW and the MMW to slab transitions, the basis can be given by 2D slab modes whose k-vector components are quantized in the y-direction by a periodic boundary condition across a slab width $w = N \cdot d$, which can be imposed as it is also satisfied by the basis of the input array of waveguides and is conserved across the transition (both the waveguide array and the slab are considered to be of infinite extent, wherein the subset of Bloch modes that satisfy this periodic boundary condition are considered in each). These are then given by $k_y = 2\pi q/w$. Within the slab, the N lowest order modes are equivalently given by $q \in [-N/2, N/2]$ and thus by $k_y \in [-\pi/d, \pi/d]$, i.e., their $k_y$ falls within the first Brillouin zone of the waveguide array. Higher order beams generated by the OPA, on the other hand, are offset from the center beam by $\Delta k_y = 2\pi/d$ and are outside the first Brillouin zone. The side-lobe suppression problem can thus be reformulated into only exciting slab modes that fall within the first Brillouin zone and are thus the N lowest modes of the slab. Since the waveguide array supports N super-modes, this in turn boils down to convert the N modes of the waveguide array into the N lowest modes of the slab, which can be accomplished by ensuring that all the transitions are adiabatic, or as close to it as possible. This can be accomplished by slowly turning on the coupling between the waveguides as they are progressively merged into a slab. The reader may refer to the literature for a more complete and rigorous analytical derivation [57, 58].

In F-OPA2, waveguides are initially single-mode and isolated, resulting in negligible mutual coupling at the input. As the waveguides are widened and transition into a uniform medium, coupling gradually increases, corresponding at first to an adiabatic transition. However, due to the fabrication constraints, once a waveguide-to-waveguide gap of 0.4 μm is reached the tapering stops, and an abrupt transition to a slab follows. This abrupt transition causes substantial power to be transferred to higher order slab modes outside of the first Brillouin zone and to the excitation of the side-lobes. Therefore, a structure needs to be introduced that enables progressive increase of the waveguide-to-waveguide coupling to a higher level while staying within the constraints of the fabrication process.

To achieve this, we leverage the double-stripe waveguide structure of the LioniX SiN platform (Fig. 2). Unlike previous designs, in which fully etched arrays had a minimum gap of 0.4 μm, here we add an intermediate transition region before the slab (89.4 μm cumulative width), in which only the top SiN layer is etched to define the waveguide array (Fig. 8(a)). This approach reduces the effective refractive index contrast between the core and cladding regions, enlarging the extent of the evanescent fields, and significantly increases the maximum coupling strength that can be achieved between the waveguides. This is compounded by the fact that owing to the lower etch depth the backfilling with $SiO_2$ becomes easier and the minimum gap between the waveguides can be reduced from 0.4 μm to 0.25 μm in this region while maintaining the 1.2 μm pitch, further increasing the coupling strength. Furthermore, to prevent super-mode back-reflections at the lateral boundaries of the waveguide array or slab, four dummy waveguides are added on each side of the array in addition to the 64 central waveguides that are supplied with light. In this structure, referred to as F-OPA3, the SMWs first transition from the fully etched to the partially etched region, after which they are tapered up over a 70 μm length from the initial SMW width to d-0.25 μm = 0.95 μm. As previously, the array incorporates a built-in parabolic phase shift with $\psi_M$ set to 45 radians.

The increased coupling between the waveguides is evidenced by Fig. 8(c), that shows the near field Poynting vector component $S_z(x, y)$ recorded after launching light in a single waveguide, in which it is apparent that the extent of the field profile in the transverse y-direction is significantly wider than for F-OPA2 (Fig. 7(d)).

The simulation results shown in Figs. 8(d) and 8(e) indicate that at a focal distance of 185 μm measured along the $u_z$-axis from the chip surface, the main lobe has longitudinal, transverse, and axial FWHM values of 8.87 μm, 1.176 μm, and 23.1 μm, in line with what has been simulated for F-OPA1 and F-OPA2. However, in F-OPA3 approximately 63% of the power is concentrated in the main lobe, with each side-lobe holding around 11.5% of the total emitted power at a spacing of 86 μm from the main one. Consequently, F-OPA3 demonstrates a substantial performance improvement compared to both F-OPA1 and F-OPA2.

Further improvement of the coupling between the waveguides requires further expansion of their modes in the transition region. As a final step in the design optimization, we now switch to an inverse taper configuration to further expand the evanescent fields. A taper with a single nano-tip waveguide is a well-known solution for mode expansion and has been widely used in edge coupler designs [59]. However, such a tip expands the field isotropically in all directions, including along the surface normal of the chip, which makes low loss transitioning to a highly confined slab mode difficult. Multi-tip or trident structures have been utilized in edge coupler devices to achieve low-loss coupling with anisotropically expanded fields [60, 61]. Based on these structures, we implement a trident-waveguide array in F-OPA4, as shown in Fig. 8(b).

SMWs with a 1.2 μm pitch are first tapered down to a 0.12 μm width over a 15 μm inverse taper length after transitioning into the partially etched region. Two additional nano-tip waveguides of identical width are then symmetrically introduced on both sides of the central one and separated by a 0.25 μm gap from it, thus forming a trident-shaped waveguide arrangement. This trident structure supports a ground and a first order mode. However, since the first order mode has the opposite symmetry than the incoming ground mode of the tapered down interconnect waveguide, only its ground mode can be excited, even though the auxiliary waveguide-tips are abruptly introduced.

At this point, further enhancement of the coupling has been enabled, however, if routed as is to the slab, these trident waveguides would result in high insertion losses at the waveguide-to-slab transition, as their fields are much more delocalized in the vertical direction than in the slab. To pull the fields back into the slab layers, the three tips are tapered up together over a subsequent length of 50 μm, at the end of which the tips and gaps are 150 nm and 250 nm wide, respectively, the latter representing the minimum gap size allowable in the fabrication process.

Due to the strong waveguide-to-waveguide coupling in this device, a larger number of dummy waveguides had to be implemented in the array. Consequently, 12 dummy waveguides are implemented on each side, resulting in a cumulative width of 108.6 μm. The 3D-FDTD simulation of the near field above the chip, resulting as previously from single waveguide excitation, is shown in Fig. 8(f) and features a strong field at far away waveguides, confirming the enhanced waveguide coupling. After far-field beam propagation, simulations show a very substantial suppression of the side-lobe power with only 1.6% of the emitted power in each of the side-lobes and the main lobe carrying 79.58% of the power, clearly surpassing previous designs (see Fig. 8(g) and 8(h)).

Figure 9, obtained from the 3D-FDTD simulations of the four OPAs with single waveguide excitation, compares the x-component of the Poynting vector, pointing in the direction of propagation of the light, in the center plane of the waveguide core after tapering of the waveguides and, for F-OPAs 2-4, after coupling into the slab region, and before the onset of the grates. While F-OPA1 and F-OPA2 do not feature significant coupling to neighboring waveguides, F-OPA3 shows coupling to the nearest neighbors and F-OPA4 shows substantial coupling across the simulated array of 25 waveguides.

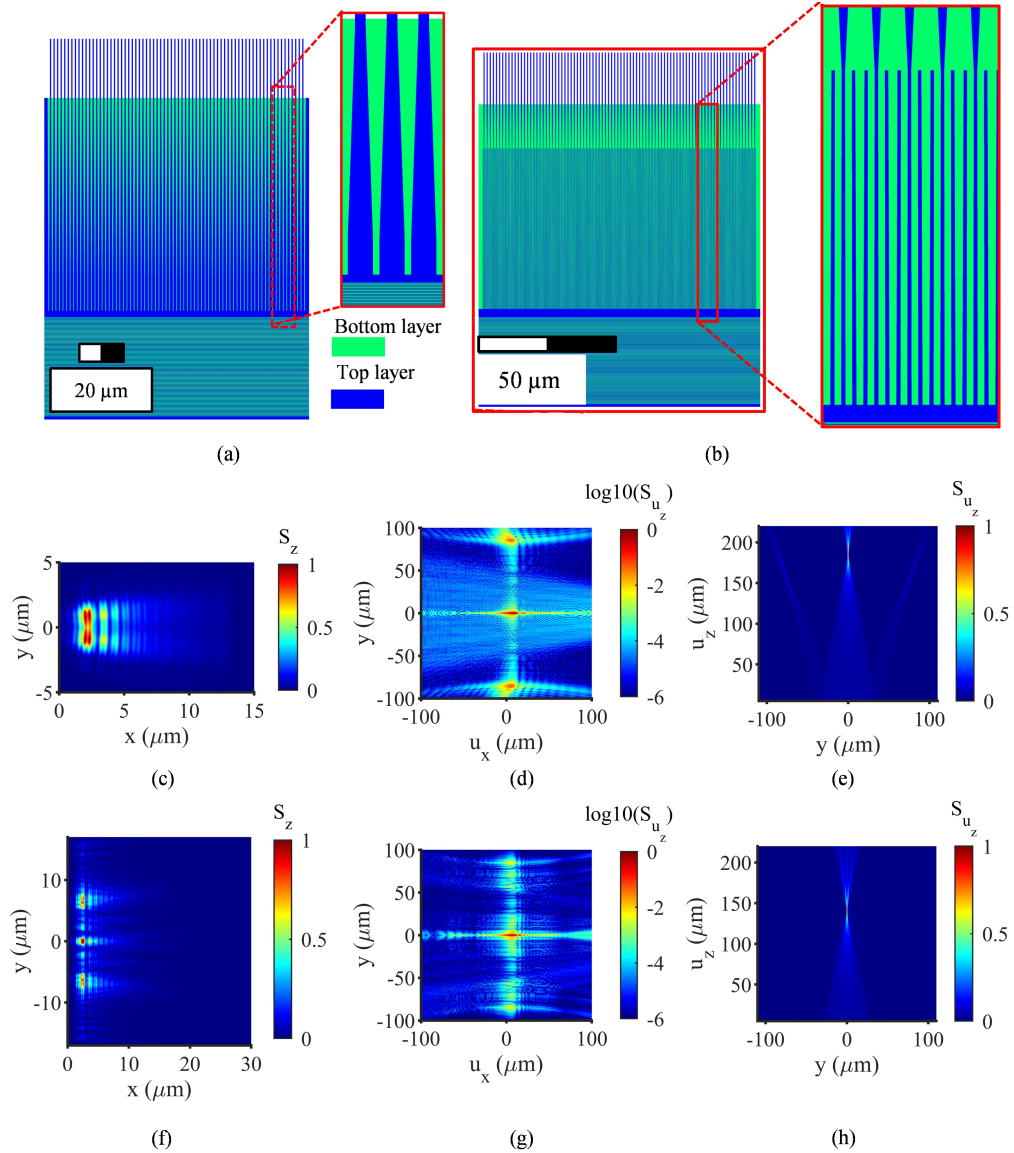

Fig. 8. OPA with strong waveguide coupling for side-lobe suppression. Layout of (a) F-OPA3 and (b) F-OPA4. Near-field profile of $S_z(x, y)$ from 3D-FDTD simulations with single-waveguide excitation in (c) F-OPA3 and (f) F-OPA4. $S_{u_z}$ in the $u_x$–y focal plane (logarithmic scale) for (d) F-OPA3 and (g) F-OPA4 and in the y–$u_z$ plane for (e) F-OPA3 and (h) F-OPA4.

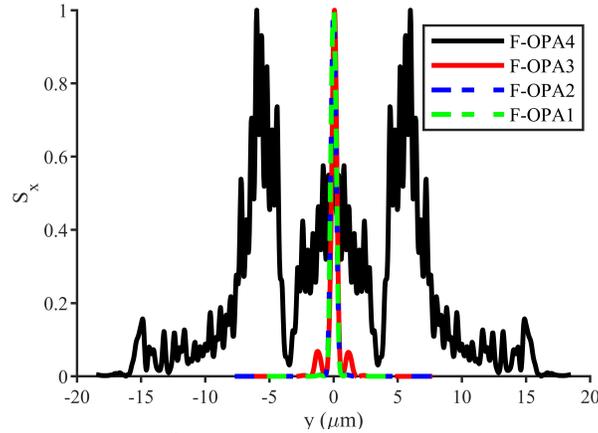

Fig. 9. Comparison of the field profiles of $S_x$ in the center plane of the waveguide core across the transverse direction for the four F-OPA designs. Coupling to the nearest neighbors is visible for F-OPA3 and significant coupling across the array is seen for F-OPA4.

Among the F-OPA designs described above, F-OPA3 was fabricated (Fig. 10(a)) and characterized using the setup depicted in Fig. 4(a). Figure 10(b) presents the measured intensity in the $u_x$–y plane around the focal spot on a logarithmic scale with a maximum intensity normalized to 0 dB. After correction of measured beam sizes, as explained in Appendix A.3, the FWHM of the focal spot is extracted to be 1.27 µm along the transverse (y-)axis, 7.07 µm along the longitudinal ($u_x$)-axis, and 21.89 µm along the axial ($u_z$)-axis. Given the resolution of the beam profiler (0.922 µm per pixel) and the step size of the motorized stage in the z-direction (5 µm), the measurements show good agreement with the simulations (1.176 µm, 8.87 µm and 23.1 µm in the transverse, longitudinal and axial directions). Notably, side-lobes are not clearly visible in the experimental data due to their low power compared to the main lobe, making them indistinguishable from scattered background light resulting from the finite waveguide coherence length (see Appendix A.1). Figure 10(c) shows the intensity along the y–$u_z$ plane with a focal spot located at $u_z$ = 200 µm.

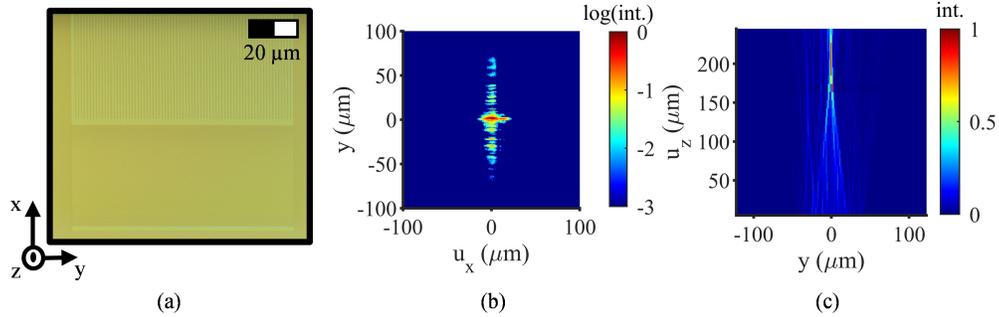

Fig. 10. Experimental data for the 64-channel F-OPA3. (a) Micrograph of the waveguide transition and the GE structure. (b) Intensity distribution recorded around the focal spot in the $u_x$–y plane, plotted on a logarithmic scale. (c) Intensity distribution in the y–$u_z$ plane.

### 3.3 Beam steering with thermo-optic phased shifters

#### 3.3.1 Design and simulation

Active phase manipulation of the OPA waveguide channels is essential for beam steering in the transverse direction or dynamic focusing along the emission axis. This manipulation can be realized using thermo-optic phase shifters via the metal layer stack provided in the LioniX

International platform. In the heater structures, a suitably increased sheet resistance is obtained by removing the gold layer, leaving the underlaying metal adhesion layers behind. In conventional designs, each channel typically requires its own phase shifter, increasing system complexity and device footprint due to the need to minimize thermal cross-talk between channels [62]. A thermal phase shifter controlled by a single input is much preferable when the phase shifts applied to the different channels have a fixed relation to each other, as this greatly facilitates routing and control. Previously in [56], we introduced meandered heater structures enabling dynamic beam steering with simplified control, similar to an approach that has been used in the silicon-on-insulator platform at infrared wavelengths [44, 63].

As shown in Fig. 11(a), inset 11(b), a heater is implemented with a meandered layout such that it overlays each waveguide of the interconnect array by an amount that scales with the channel index p, thus introducing a linear phase shift across the array with an increment that depends on a single current sent through the heater. This concept has been extended to allow steering of the beam in both transverse directions, by implementing two meandered heaters in complementary push-pull configuration, wherein the overlay of the second meandered heater with the waveguides scales with N-p instead. Since they take complementary triangular shapes, they can be stacked vertically, reducing the cumulative waveguide length and thus requirements on the waveguide coherence length (see Appendix A.1). The structure has been further duplicated with two heaters of the same type sharing a common control current to increase the steering range within the reliability constraints of the heater layer (in the following, simulation results or characteristics such as resistance always refer to the two cascaded meanders driven in series, when applicable).

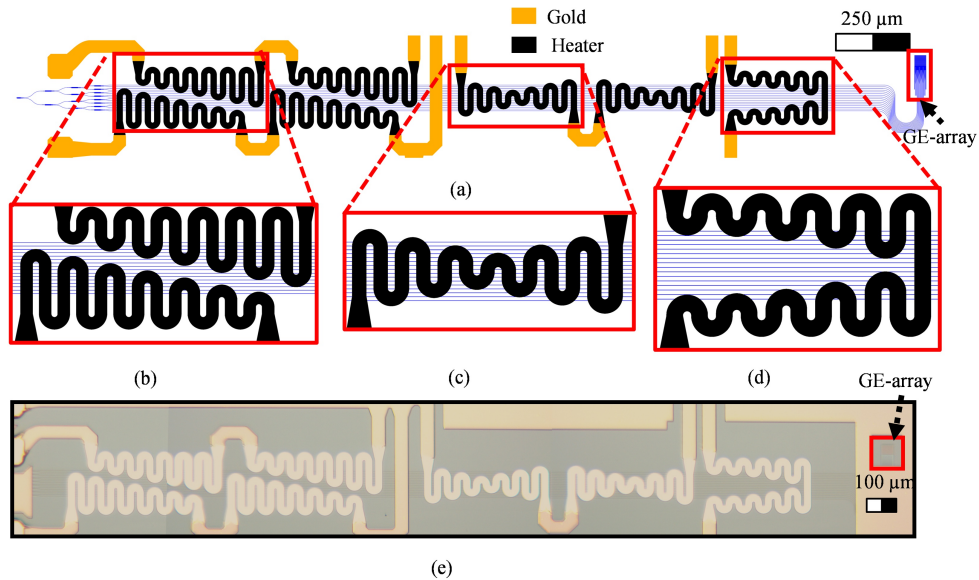

Fig. 11. (a) Layout of the 16-channel F-OPA3 integrated with linear and focusing/defocusing meandered phase shifters. (b) Linear phase shifters designed for transverse beam steering in push-pull configuration. Parabolic phase shifters for adaptive (c) focusing and (d) defocusing. (e) Micrograph of the fabricated OPA with thermo-optic phase shifters.

This concept has been further generalized to allow for dynamic focusing and defocusing of the beam, i.e., for moving the focal spot closer or further from the PIC surface, as shown in the insets 11(c) and 11(d), respectively. In this case, the overlay scales in a parabolic manner with the waveguide index, following Eq. (8). Fringe effects resulting from thermal crosstalk can be optimized by inspecting the induced phase shift in simulations and adaptively updating the positions of the meanders in the layout, to obtain the best consistency with the targeted parabolic dependency and the lowest common temperature offset (that reduces the electrical power efficiency of the device). Here too, the focusing meander has been duplicated to increase the axial steering range, while this is not the case for the defocusing one, that, however, already starts out with a longer length and thus a higher resistance (see below).

These phase shifters have been applied to the F-OPA3 configuration, as already described and characterized in the previous section, with some changes. A micrograph is shown in Fig. 11(e). As a requirement for interposing the heaters, the interconnect waveguide array has been prolonged, which results in increased requirements for the waveguide coherence length. The number of independent waveguide channels has also been reduced to 16 here, reducing the numerical aperture of the array and thus the achievable transverse spot size. The pitch between the waveguides before the GE has also been slightly increased to 1.4 µm and the maximum hard-wired parabolic phase shift $\psi_M$ is 7 radian. Other design parameters are identical to F-OPA3 as described in Section 3.2.

A 4-level MMI-based distribution network is followed by the phase shifter region, in which waveguides run in parallel over a cumulative length of 2.45 mm with a spacing of 5.5 µm. The sheet resistance of the Ta/Pt heater layer is measured to be 3.35 Ω/sq, assuming negligible resistance from the gold interconnect traces. Due to fabrication constraints, parallel heater traces have a width of 20 µm and a gap of 10 µm. Meanders have an inner bend radius of 5 µm. The resistances of the cascaded linear phase shifters, the cascaded focusing phase shifters and the defocusing phase shifter are 428.7 Ω, 376 Ω and 233.7 Ω, respectively.

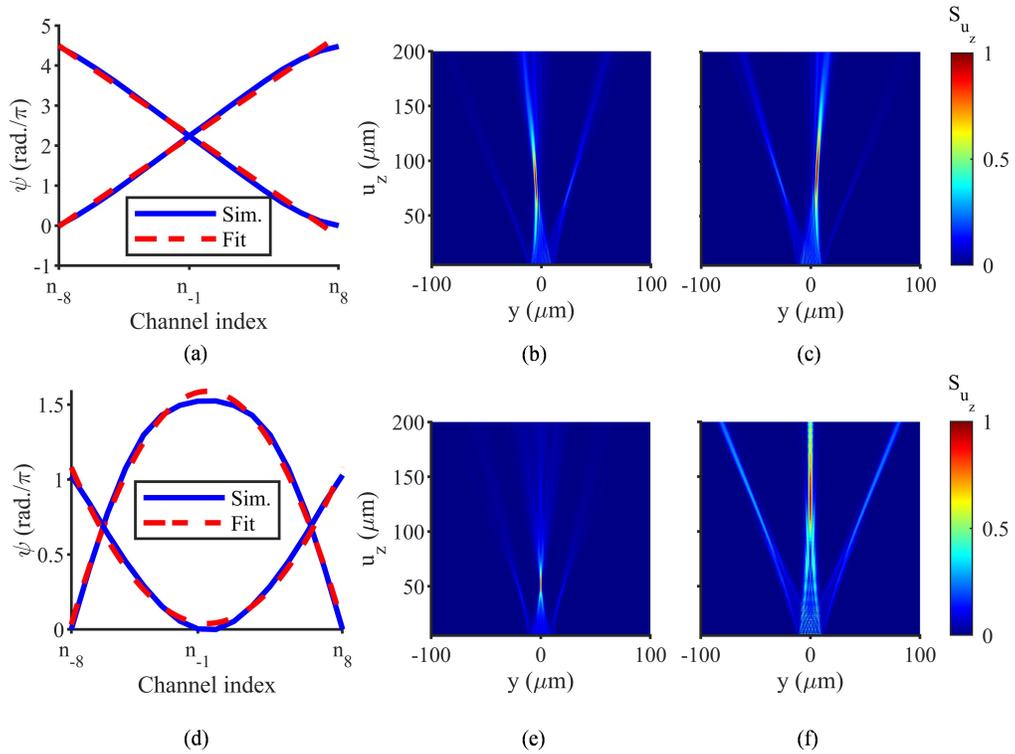

Fig. 12. Simulation results for the 16-channel OPA with integrated thermal phase shifters. (a) Phase delays applied across the waveguide array for positive and negative transverse beam steering by the triangular shaped meandered phase shifters. Field distributions ($S_{u_z}$) in the y–$u_z$ plane when applying 50 mA of electrical power for (b) negative and (c) positive transverse steering angles. (d) Phase delays applied for focusing and defocusing and distribution of $S_{u_z}$ in the y–$u_z$ plane for an applied tuning current of (e) 42 mA (focusing) and (f) 46 mA (defocusing).

3D thermal simulations were performed to analyze heat dissipation and temperature gradients along the waveguide array. Using the room temperature thermo-optic coefficients of SiN and SiO$_2$, 2.51×10$^{-5}$ K$^{-1}$ and 0.96×10$^{-5}$ K$^{-1}$, respectively [64], the relative phase gradient is calculated as

$$\psi_p = \frac{2\pi \int \Delta n_{eff_p}(l) dl}{\lambda_0} \quad (11)$$

where $\Delta n_{eff_p}(l)$ is the local effective index change in the waveguide of index p and the integral is taken over the length of the waveguide.

Simulation results are summarized in Fig. 12, with Fig. 12(a) showing the phase shifts applied across the waveguides when the triangular meanders from Fig. 11(b) are actuated with 50 mA of electrical current. It can be seen that the resulting phase shifts have close to constant increments, as targeted. They result in steering angles of ±3.41°, as shown in Fig. 12(b) and 12(c) after beam propagation. For the focusing and defocusing phase shifters, Fig. 12(d), applying 42 mA and 46 mA respectively, shifts the focal point from $u_z$ = 50.63 to 128.4 μm (Fig. 12(e) and (f)). As discussed in the next section, these steering ranges are actually underestimated compared to what is achieved experimentally.

### 3.3.1 Measurement and discussion

The 16-channel OPA was experimentally characterized with results shown in Fig. 13. For reference, Fig. 13(a) shows the intensity profile recorded in the y–$u_z$ plane when none of the phase shifters are actuated. It features a focal point at $u_z$ = 110 μm, somewhat farther from the PIC surface than expected from simulations, but this can also be due to the difficulty of determining an absolute distance in the experimental setup (relative distance changes are measured much more precisely). The beam has an initial transverse tilt of ~2.1°, likely caused by a slight gradient in the effective index across the extended waveguides caused by a spatially dependent process bias in film thickness or waveguide width, or by a slight angular misalignment between the e-beam lithography defining the GE trenches and the optical lithography defining the waveguide to slab interface. The emission profile also features pronounced stray light, which indicates that the waveguide coherence length is limiting performance after extension of the waveguide array to fit the meanders.

Applying 50 mA to one of the linear heaters results in an angle of approximately -3°, as shown in Fig. 13(b). Figures 13(d) and 13(e) feature a focal spot movement from 65 μm to 140 μm after applying 30 mA on de-/focusing phase shifters. Figures 13(c) and 13(f) compare simulative with experimental data for beam steering and dynamic focusing. In Fig. 13(f), negative and positive current levels correspond to the focusing and defocusing phase shifters, respectively, so that the data can be summarized in a single plot. Measurements feature ±5.1° of beam steering and a 204 μm movement of the focal spot along the axial direction for the current ranges investigated by simulations, which shows that the latter consistently underestimate steering ranges. This discrepancy is likely due to memory constraints limiting the simulation domain size. The silicon substrate of the PIC, that limits heat flux between the device and the underlying heat sink, is 525 μm thick. However, we could only model a 100 μm thickness in the 3D simulations before running into a memory wall. This increases the heat flux to the heat sink, modeled as a fixed temperature boundary condition, reducing the modeled efficiency of the phase shifters. Additionally, another limitation arises from the boundaries along the x- and y-axes in the simulation region, that have been selected to be insulating. While these increase the efficiency of any given single phase shifter by preventing the heat from escaping laterally out of the simulation domain, they also tend to homogenize the temperature across the waveguide array, so that the modeled steering, relying on phase differences between the waveguides, is further reduced.

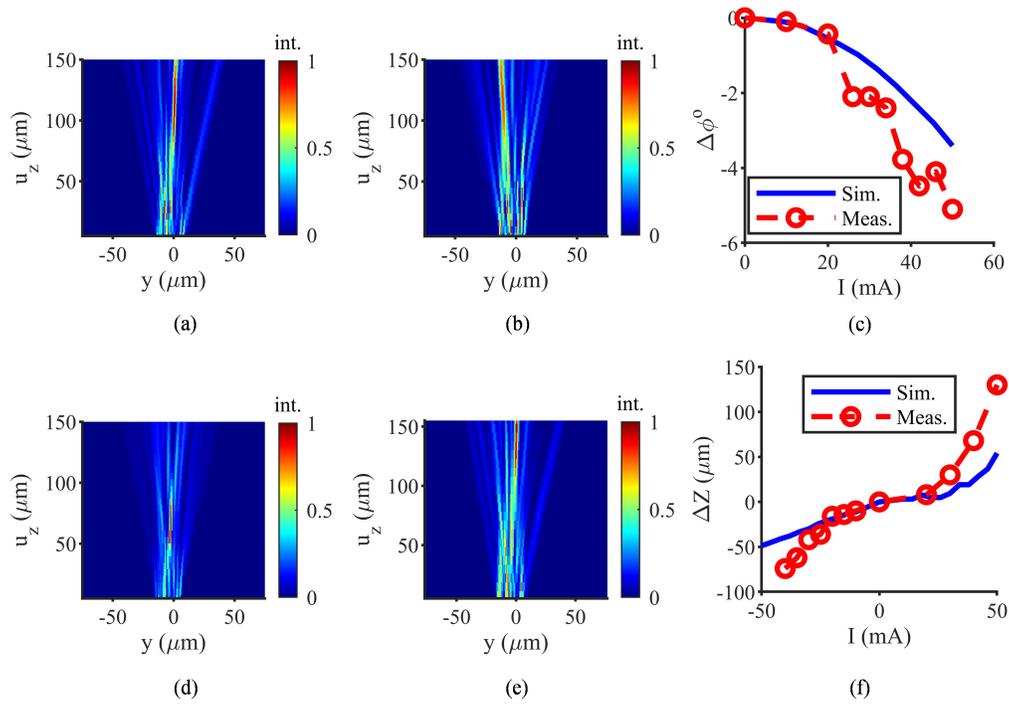

Fig. 13. Experimental characterization of the fabricated 16-channel F-OPA with meandered phase shifters. (a) Baseline $y$–$u_z$ plane intensity distribution without phase shifter actuation, showing a built-in tilt. (b) Measured intensity distribution featuring a beam steering offset of -5.1° after applying 50 mA to one of the linear phase shifters. Focal spot moving along the axial direction from (d) 65 µm to (e) 140 µm with current levels of 30 mA applied to (d) the focusing and (e) the defocusing phase shifter. Comparisons between simulation and experiment for (c) beam steering and (f) focusing/defocusing. In (f), negative currents correspond to focusing and positive ones to defocusing.

Table 1 summarizes all F-OPAs presented in this work, along with the measurement results of the fabricated devices, which demonstrate subcellular resolution along the transverse axis.

Table 1: Comparison of designed and characterized focusing optical emitters.

| Device name | Number of channels | Grating structure | Pitch (µm) | $\psi_M$ (rad.) | Main/side-lobe/ stray power ratio % | Focal spot FWHM [$\Delta u_x$, $\Delta y$, $\Delta u_z$] (µm) | Focal distance, $u_z$ (µm) |
|---|---|---|---|---|---|---|---|
| FGE | 1 | GE1 | — | — | — | [11.86, 1.44, 35.52] | 232.5 |
| **Measurement** | | | | | — | **[9.36, 1.45, 33.4]** | **211** |
| F-OPA1 | 64 | GE2 | 1.2 | 45 | 53.78 / 16.3 / 13.62 | [11, 1.16, 23] | 192.3 |
| F-OPA2 | 64 | GE2 | 1.2 | 45 | 56.65 / 14.10 / 15.15 | [9.68, 1.17, 23.06] | 187.1 |
| F-OPA3 | 64 | GE2 | 1.2 | 45 | 63 / 11.5 / 14 | [8.87, 1.18, 23.1] | 185 |
| **Measurement** | | | | | **59.84 / —** | **[7.07, 1.27, 21.89]** | **200** |
| F-OPA4 | 64 | GE2 | 1.2 | 45 | 79.58 / 1.6 / 17.22 | [7.91, 1.13, 21.23] | 141.3 |
| F-OPA3 | 16 | GE2 | 1.4 | 7 | 68.3 / 11.66 / 8.38 | [6.36, 1.8, 53] | 83 |
| **Measurement** | | | | | **38.17 / —** | **[6.36, 1.65, 34.38]** | **110** |

For the 64-channel F-OPA3, the main lobe power ratio from simulations (63%) aligns well with the measurement (59.84%). However, there is a significant discrepancy for the 16-channel F-OPA3, with the main lobe power ratio modeled as 68.3% and measured as 38.2%. This is a consequence of the increased waveguide lengths and the finite coherence length of the waveguides, which result in a substantial portion of the light being radiated as stray light outside of the main beams (see appendix A.1). Since in both F-OPA devices the side-lobes cannot be clearly identified in the experimental data, only the power in the main lobe is extracted for both experimental datasets.

As an outlook for further work, focusing and high-resolution excitation could also be achieved along the longitudinal axis with GEs with apodization of the grating teeth along the direction of propagation, similar to the approach described in [23]. However, this poses additional challenges for fabrication due to the required minimum feature size, particularly if a steep emission angle is to be maintained at visible wavelengths, that requires a small baseline pitch. Furthermore, while this work utilizes a one-photon excitation method, higher axial resolution can be achieved using two-photon excitation [65].

## 4. Conclusion

We have described a MEA integrated into an SiN PIC capable of performing *ex-vivo* retina characterization. The employed PIC contains optical emitters with focusing capability in one direction, achieving a sub-cellular focal spot in the retinal layers on top of the chip. The grating structures that serve as the emitter units on this PIC provide beam steering of 4.26° in the longitudinal direction when applying wavelengths within a range of 525 ± 15 nm. The first type of emitter investigated is a curved grating structure, producing a 1.45 μm focal spot in the transverse direction. Subsequently, focusing OPAs were explored because of their beam steering and transverse focusing and defocusing capabilities. We studied four different F-OPA designs aiming at suppressing side-lobes while focusing the beam, which is achieved by a combination of built-in and dynamic phase shifts. As a proof of concept, one of the fabricated F-OPA devices demonstrated ~60% of the power concentrated in the main lobe, with each side-lobe containing about 11.5% of the power, while achieving a transverse focal spot of 1.27 μm. This was achieved by increasing the coupling between waveguides in a transition region interposed before the slab. To improve on this, we introduced a new F-OPA design based on trident waveguides with only 1.6% of the power in each side-lobe and nearly 80% in the main lobe. Additionally, meandered thermo-optic phase shifters integrated onto the F-OPAs enable dynamic beam steering and axial focal spot translation with a single electrical input per steering direction, thus reducing the complexity and footprint of the overall device. A transverse steering angle of ±5.1° and a focal spot translation of 204 μm were experimentally achieved using these tuners with currents below 50 mA.

## Appendix:

*A.1 Coherent length analysis*

Even though the meandered thermo-optic phase shifters provide a compact footprint compared to conventional methods, integration of transverse steering and axial focusing/defocusing phase shifters requires relatively long waveguides. In the device presented in Section 3.3, the total waveguide length is almost 3 mm per channel, resulting in the finite coherence length of the waveguides limiting the quality of the beam shaping in the form of stray light (see Fig. 13). To assess the effect of the finite waveguide coherence length, we simulate the device under the assumption of varying coherence length and record the amount of stray light defined as the total power that does not lie within the lobe windows described in Section 3.3.1. In the reference case, in which the coherence length is assumed to be infinite, the stray light only makes up for 8.38% of the total. Phase errors $\delta\psi$ are introduced independently for each channel by sampling a zero-mean normal distribution with a variance $2L/L_{coh}$, where L is the physical waveguide length and $L_{coh}$ is the assumed coherence length [66], i.e.,

$$\psi'_p = \psi_p + \delta\psi_p, \text{with } \delta\psi \sim \mathcal{N}\left(0, \sqrt{\frac{2L}{L_{coh}}}\right) \quad (12)$$

where $\psi_p$ is the ideal phase setting from the reference case. $L_{coh}$ is varied between 2 mm and 13 mm. The results, summarized in Fig. 14, show how increasing phase errors resulting from a reduced coherence length lead to an increase of the percentage of stray light. The "C.L. sim." represents this trend, while the "Inf. C.L." level shows a performance benchmark corresponding to an infinite coherence length. The dashed red curve represents the stray light ratio obtained from the measurements which is calculated from the stray light window defined in infinite coherence length simulation. The dashed black line shows a linear regression applied to the simulation data and represents an average performance as a function of coherence length. After subtracting it from the blue curve, we extract the standard deviation of the residue. The green curves represent the linear regression offset by plus/minus one standard deviation. This analysis highlights the critical role the coherence length plays in determining the beam quality of the SiN-PIC-based OPA. Based on this data, we may expect the coherence length to be in the order of 7 mm, but it should be noted that there is a very large spread in the simulation data due to the stochasticity of the phase error sampling, so that this order of magnitude is only indicative here.

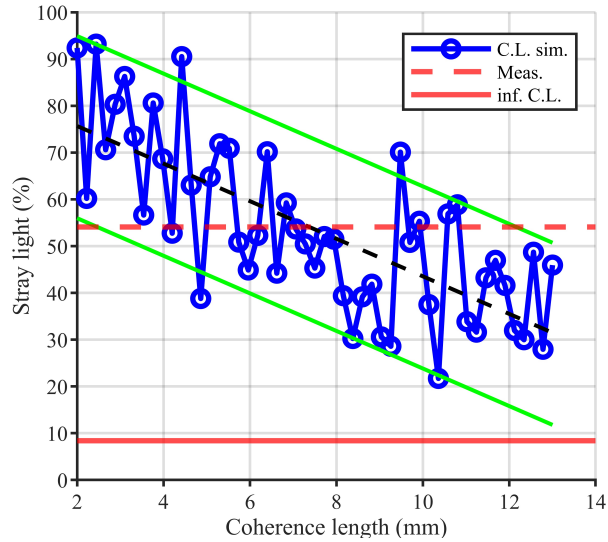

Fig. 14. Coherence length analysis for the 16-channel OPA described in Section 3.3. Coherence lengths in the range of 2 to 13 mm are considered. The solid and dashed red lines show the simulated performance for infinite coherence length and as extracted from the experimental data, respectively. The black dashed curve shows a linear regression of the simulation data (blue curve), with the plus/minus one standard deviation confidence interval shown by green curves.

*A.2 Gaussian fit*

In the main text, the power of each lobe emitted by the OPA is quantified by integrating the emitted power within a total width of two MFDs extracted from a Gaussian fit of the same lobe (with the MFD defined as the distance between the points where the power drops to $1/e^2$ of its maximum).

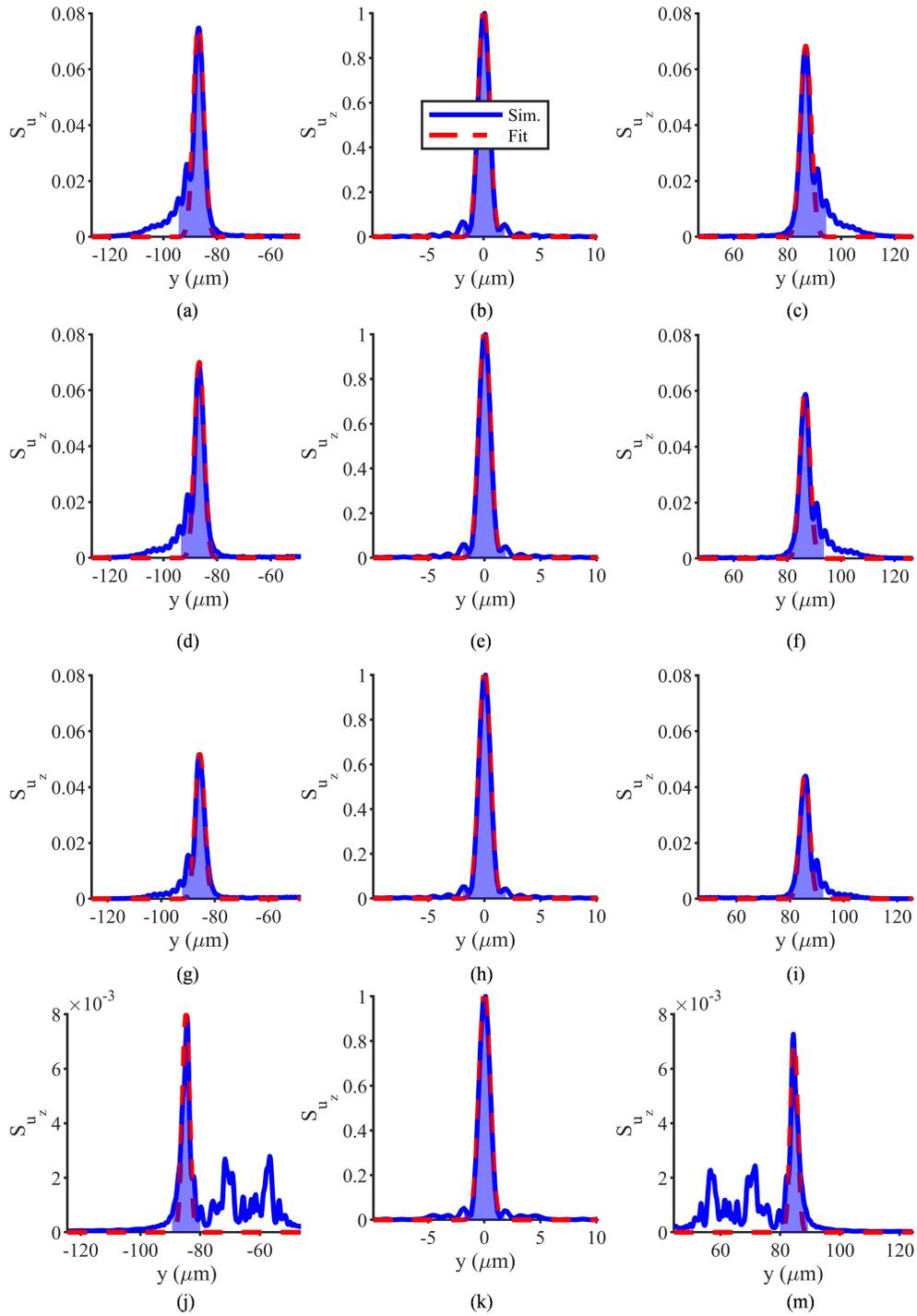

Fig. 15: Gaussian fits of the Poynting vector distribution $S_{u_z}(y)$ along the y-axis for F-OPAs 1-4 at the focal spot. Left and right columns show the side-lobes, while the center one shows the main lobe normalized to a unit peak intensity. (a-c) F-OPA1, (d-f) F-OPA2, (g-i) F-OPA3, and (j-m) F-OPA4. The gray-blue region shows 2×MFD intervals of the Gaussian fits in which the power of each lobe is integrated.

However, as shown in Fig. 15, a certain amount of power in the lobes does not lie in this range due to their non-Gaussian shape, specifically in F-OPA4. This power is considered as part of the stray light, leading to higher stray power in F-OPA4 compared to the other devices, as shown in table 1. If the entire power surrounding the side-lobes of F-OPA4 as shown in Fig. 15, panels (j) and (m), was allocated to them, their respective share of the total power would increase to an average 3.85% per side-lobe, which still remains very small. It can thus be seen that the overall conclusions on the effectivity of the side-lobe suppression scheme are not sensitive on the exact methodology followed to evaluate their power, but are rather general.

### A.3 Correction of focal spot measurement

The focal spot dimensions are obtained from fitting a Gaussian on the simulation and measurement data, as illustrated in the previous section A.2, and calculating the FWHM of the fit. The experimental data is, however, limited by the resolution of the objective embedded in the beam profiler, see Fig. 4(a) for a schematic of the setup. The lens's clear aperture diameter is $D_{eff} \approx 3.8$ mm and its focal length $f = 8$ mm with a magnification of $m = 4$. This results in an object distance $u = f(1 + m)/m = 10$ mm. The corresponding numerical aperture (NA) is estimated as $NA \approx D_{eff}/2u = 0.19$. Using Abbe's diffraction limit, the FWHM of the corresponding Airy disk that corresponds to the blur of the imaging optics is calculated as $FWHM_b = 0.51\lambda_0/NA = 1.4$ µm, with $\lambda_0$ the free space wavelength. Consequently, the measured transverse FWHM of the focal spot (along the y-axis) is expected to be obtained by convolving the simulated Gaussian field with the resolution of the instrument, resulting in

$$FWHM_{ex}^2 = FWHM_{sim}^2 + FWHM_b^2$$

Considering beam divergence in only one direction (transverse) and collimation along the longitudinal direction, due to the much wider beam width in that direction, the FWHM of the beam along the axial direction no longer corresponds to two times the textbook expression of the Rayleigh length $l_R$ for isotropic beams. It is given by an expression that is a factor $\sqrt{3}$ larger, so that the beam width along y, that scales as $\sqrt{1 + (u_z/l_R)^2}$, is doubled at the edges of $FWHM_{u_z}$:

$$FWHM_{u_z} = \sqrt{3}\pi FWHM_y^2/\ln(2)\lambda_0$$

The following table shows the details of the FWHM extraction from experimental data for both the transverse (y) and axial ($u_z$) directions for the three measured emitters. The expected values are obtained from the simulation results taking the resolution limit of the imager into account and can be compared with the measurement results. The corrected values are obtained from the measurements after taking the resolution limit into account and can be compared with the simulations. Corrected values for the FWHM along $u_z$ are rescaled according to $FWHM_y^2$ in the last column. The only significant discrepancy between simulations and measurements can be seen to be for the FWHM along $u_z$ for the 16-channel device, which we attribute to the corresponding experimental data suffering from a high level of stray light due to the finite waveguide coherence length, which makes this number particularly difficult to extract.

Table 2: Focal spot size analysis for the fabricated focusing emitters.

| | **FWHM Transverse (µm) - $FWHM_y$** | | | |
| --- | --- | --- | --- | --- |
| | **Simulation** | **Expected measurement** | **Measurement** | **Corrected measurement** |
| **FGE** | 1.44 | 2.01 | 2.02 | 1.46 |
| **F-OPA3, 64 ch.** | 1.18 | 1.83 | 1.89 | 1.27 |
| **F-OPA3, 16 ch.** | 1.80 | 2.28 | 2.17 | 1.66 |

|  | FWHM Axial (µm) - $FWHM_{u_z}$ | | | |
| --- | --- | --- | --- | --- |
|  | Simulation | Expected measurement | Measurement | Corrected measurement |
| FGE | 35.5 | 60.4 | 64.4 | 33.6 |
| F-OPA3, 64 ch. | 23.1 | 50.1 | 48.5 | 21.9 |
| F-OPA3, 16 ch. | 53.0 | 77.7 | 58.9 | 34.5 |

**Funding.** German Research Foundation (DFG, RTG 2610).

**Disclosures.** The authors declare no conflicts of interest.

**Data availability.** Data underlying the results presented in this paper are not publicly available at this time but may be obtained from the authors upon reasonable request.